\documentclass[twocolumn,showpacs,preprintnumbers,amsmath,amssymb,nofootinbib]{revtex4}
\usepackage{amsmath}
\usepackage{graphicx}
\usepackage{calc}
\usepackage{bm}
\usepackage{float}

\begin{document}
\newcommand{\beq}{\begin{equation}}
\newcommand{\eeq}{\end{equation}}

\title{Band Flattening and Landau Level Merging in Strongly-Correlated Two-Dimensional Electron Systems}
\author{V.~T. Dolgopolov$^a$, M.~Yu.~Melnikov$^a$, A.~A. Shashkin$^a$ and S. V. Kravchenko$^b$\vspace{2mm}}
\affiliation{$^a$Institute of Solid State Physics, Chernogolovka, Moscow District, 142432, Russia\\
$^b$Physics Department, Northeastern University, Boston, Massachusetts 02115, USA}

\begin{abstract}
We review recent experimental results indicating the band flattening and Landau level merging at the chemical potential in strongly-correlated two-dimensional (2D) electron systems. In ultra-clean, strongly interacting 2D electron system in SiGe/Si/SiGe quantum wells, the effective electron mass at the Fermi level monotonically increases in the entire range of electron densities, while the energy-averaged mass saturates at low densities. The qualitatively different behavior of the two masses reveals a precursor to the interaction-induced single-particle spectrum flattening at the chemical potential in this electron system, in which case the fermion ``condensation'' at the Fermi level occurs in a range of momenta, unlike the condensation of bosons. In strong magnetic fields, perpendicular to the 2D electron layer, a similar effect of different fillings of quantum levels at the chemical potential --- the merging of the spin- and valley-split Landau levels at the chemical potential --- is observed in Si inversion layers and bilayer 2D electron system in GaAs. Indication of merging of the quantum levels of composite fermions with different valley indices is also reported in ultra-clean SiGe/Si/SiGe quantum wells.
\end{abstract}
\maketitle

\date{\today}

\section{Introduction}

In a non-interacting fermion system with a continuous spectrum, the occupation probability for a quantum state at fixed chemical potential and temperature is a function of the single-particle energy only \cite{landau1980statistical}. If the temperature tends to zero, the energy interval separating the filled and empty quantum states also tends to zero. For free particles, there appears a Fermi surface in momentum space with dimensionality $d-1$, where $d$ is the dimensionality of the fermions.

In general, this reasoning is not true for interacting fermions \cite{khodel1990superfluidity,volovik1991a,nozieres1992properties,khodel2008topology,shaginyan2010scaling,clark2012strongly,zverev2012microscopic}. In this case the single-particle energy depends on electron distributions, and the occupation numbers of quantum states at the chemical potential can be different, falling within the range between zero and one. A topological phase transition has been predicted at $T=0$ in strongly correlated Fermi systems that is related to the emergence of a flat portion of the single-particle spectrum at the chemical potential as the strength of fermion-fermion interaction is increased (the top inset of Fig.~\ref{fig1}). This transition is associated with the band flattening or swelling of the Fermi surface in momentum space, which is preceded by an increasing quasiparticle effective mass $m_{\text F}$ at the Fermi level that diverges at the quantum critical point. The creation and investigation of flat-band materials is currently a forefront area of modern physics \cite{heikkila2011flat,bennemann2013novel,peotta2015superfluidity,volovik2015from}. The interest is ignited, in particular, by the fact that, due to the anomalous density of states, the flattening of the band may be important for the construction of room temperature superconductivity. The appearance of a flat band is theoretically predicted \cite{amusia2015theory,camjayi2008coulomb,yudin2014fermi} in a number of systems, including heavy fermions, high-temperature superconducting materials, $^3$He, and two-dimensional electron systems.

The role of electron-electron interactions in the behavior of two-dimensional electron systems increases as the electron density is decreased. The interaction strength is characterized by the Wigner-Seitz radius, $r_{\text s}=1/(\pi n_{\text s})^{1/2}a_{\text B}$ (here $n_{\text s}$ is the electron density and $a_{\text B}$ is the effective Bohr radius in semiconductor), which in the single-valley case is equal to the ratio of the Coulomb and kinetic energies.

It has been experimentally shown that with decreasing electron density (or increasing interaction strength) in ultraclean SiGe/Si/SiGe quantum wells, the mass at the Fermi level monotonically increases in the entire range of electron densities \cite{melnikov2017indication}. In contrast, the energy-averaged mass saturates at low densities. The qualitatively different behavior of the two masses reveals a precursor to the interaction-induced single-particle spectrum flattening at the Fermi level in this electron system.

For an interacting fermion system placed in strong perpendicular magnetic fields, one expects a similar effect of different fillings of quantum levels at the chemical potential. Given the energies of two quantum levels intersect each other when varying an external parameter, these can be the same as the chemical potential over a range of parameter values, \textit{i.e.}, the levels can merge at the chemical potential over this range \cite{khodel2007merging}. The level merging implies that there is an attraction between two partially-filled quantum levels. The merging interval is determined by the possibility of redistributing quasiparticles between the levels. The effect of merging is in contrast to a simple crossing of quantum levels at some electron density/magnetic field value. Experimentally, such merging of Landau levels has been detected in Si metal-oxide-semiconductor field-effect transistors (MOSFETs) \cite{shashkin2014merging} and GaAs-based bilayer structures \cite{shashkin2015interaction}. Furthermore, in ultra-clean SiGe/Si/SiGe quantum wells, an indication of merging of the composite fermion levels with different valley indices has been reported \cite{dolgopolov2021valley}. Below we review pertaining recent experimental data.

\section{Band flattening at the Fermi level}

We start with an indication of band flattening at the Fermi level reported in Ref.~\cite{melnikov2017indication}. Raw experimental data obtained in strongly correlated 2D electron systems can be divided into two groups: (i)~data describing the electron system as a whole, like the magnetic field required to fully polarize electron spins, the thermodynamic density of states, or magnetization of the electron system, and (ii)~data related solely to the electrons at the Fermi level, like the amplitude of the Shubnikov-de~Haas oscillations yielding the effective mass $m_{\text F}$ and Land\'e $g$-factor $g_{\text F}$ at the Fermi level. As a rule, the data in the first group are interpreted using the quasiparticle language in which the energy-averaged values of effective mass, $m$, and Land\'e $g$-factor, $g$, are used. To determine the values, the formulas that hold for the case of non-interacting electrons are employed. Although this approach is ideologically incorrect, the results for $m$ and $g$ often turn out to be the same as the results for $m_{\text F}$ and $g_{\text F}$. Particularly, in a 2D electron system in Si MOSFETs, simultaneous increase of the energy-averaged effective mass and that at the Fermi level was reported in earlier publications \cite{kravchenko2004metal,shashkin2005metal,pudalov2006metal,shashkin2002sharp,mokashi2012critical,dolgopolov2015two,kuntsevich2015strongly}; it was found that the effective mass is strongly enhanced at low densities while the $g$-factor stays close to its value in bulk silicon, which did not confirm the occurrence of the Stoner instability in a 2D electron system in silicon. The mass renormalization is independent of disorder, being determined by electron-electron interactions only \cite{shashkin2007strongly}. The strongly enhanced effective mass in Si MOSFETs was interpreted in favor of the formation of the Wigner crystal or a preceding intermediate phase whose origin and existence can depend on the level of disorder in the electron system.

The experimental results reported in this section were obtained in ultra-low-disordered (001) SiGe/Si/SiGe quantum wells described in detail in Refs.~\cite{melnikov2015ultra, melnikov2017unusual}. The maximum electron mobility in these samples reached 240~m$^2$/Vs, which is the highest mobility reported for this electron system and is some two orders of magnitude higher than the maximum electron mobility in the least disordered Si MOSFETs. The parallel-field magnetoresistance (\textit{i.e.}, magnetoresistance measured in the configuration where the magnetic field is parallel to the 2D plane) allows one to determine the field of the full spin polarization, $B_{\text c}$, that corresponds to a distinct ``knee'' of the experimental dependences followed by the saturation of the resistance \cite{okamoto1999spin,vitkalov2000small} (see the bottom inset to Fig.~\ref{fig2}). The magnetic field where the spin polarization becomes complete is plotted as a function of electron density in Fig.~\ref{fig2} for two samples. Over the electron density range $0.7\times 10^{15}$~m$^{-2}<n_{\text s}<2\times 10^{15}$~m$^{-2}$, the data are described well by a linear dependence that extrapolates to zero at $n_{\text s}\approx 0.14\times 10^{15}$~m$^{-2}$ (dashed black line). However, at lower electron densities down to $n_{\text s}\approx 0.2\times 10^{15}$~m$^{-2}$ (up to $r_{\text s}\approx 12$), the experimental dependence $B_{\text c}(n_{\text s})$ deviates from the straight line and linearly extrapolates to the origin.

The solid red line in Fig.~\ref{fig2} shows the polarization field $B_{\text c}(n_{\text s})$ calculated using the quantum Monte Carlo method \cite{fleury2010energy}. The experimental results are in good agreement with the theoretical calculations for the clean limit $k_{\text F}l\gg 1$ (here $k_{\text F}$ is the Fermi wavevector and $l$ is the mean free path), assuming that the Land\'e $g$-factor, renormalized by electron-electron interactions, is equal to 2.4. Although in Ref.~\cite{fleury2010energy} Land\'e $g$-factor was equal to 2, the reason for the 20\% discrepancy between the theory and experiment may be due to the finite size of the electron wave function in the direction perpendicular to the interface. Besides, the product $k_{\text F}l$ decreases with decreasing electron density, which leads to a downward deviation in the theoretical dependence, as shown by the dotted red line in the upper inset to Fig.~\ref{fig2}.

\begin{figure}
\centering
\includegraphics[width=0.9\linewidth]{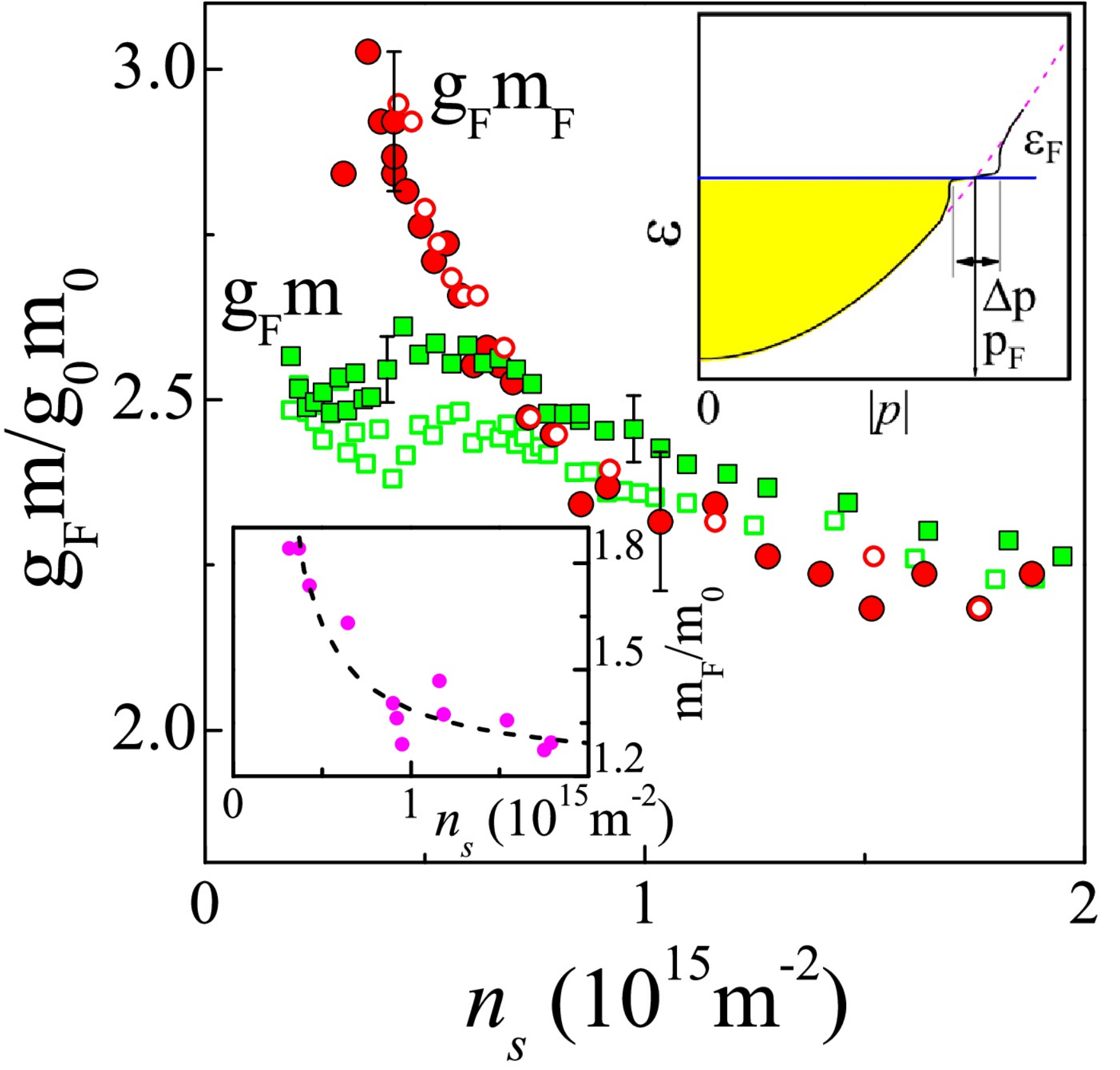}
\caption{Product of the Land\'e factor and effective electron mass in SiGe/Si/SiGe quantum wells as a function of electron density determined by measurements of the field of full spin polarization (squares) and Shubnikov-de Haas oscillations (circles) at $T\approx 30$~mK. The empty and filled symbols correspond to two samples. The experimental uncertainty corresponds to the data dispersion and is about 2\% for the squares and about 4\% for the circles ($g_0=2$ and $m_0=0.19\,m_{\text e}$ are the values for noninteracting electrons). The top inset shows schematically the single-particle spectrum of the electron system in a state preceding the band flattening at the Fermi level (solid black line). The dashed violet line corresponds to an ordinary parabolic spectrum. The occupied electron states at $T=0$ are indicated by the shaded area. Bottom inset: the effective mass $m_{\text F}$ versus electron density determined by analysis of the temperature dependence of the amplitude of Shubnikov-de Haas oscillations, similar to Ref.~\cite{melnikov2014effective}. The dashed line is a guide to the eye.  From Ref.~\cite{melnikov2017indication}.}
\label{fig1}
\end{figure}

\begin{figure}
\centering
\includegraphics[width=.855\linewidth]{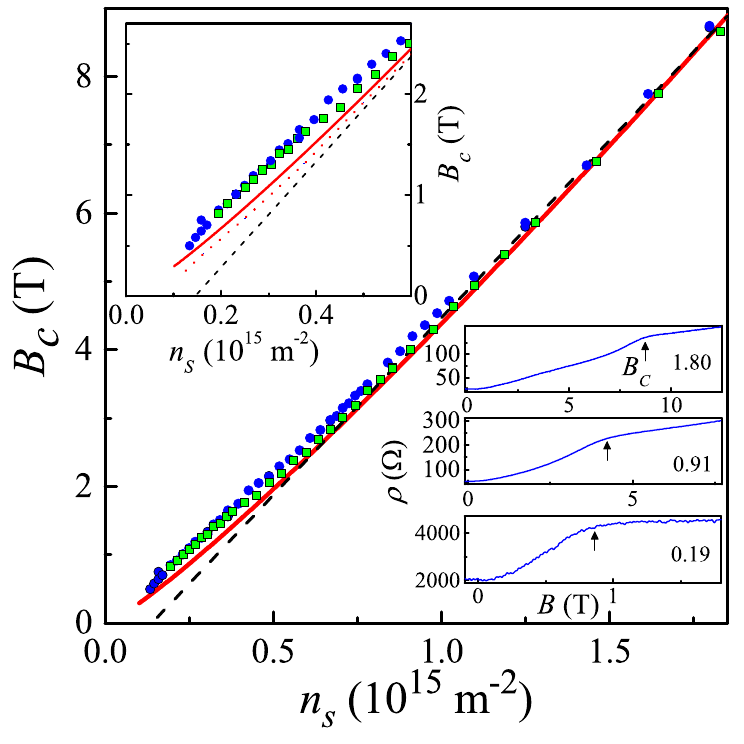}
\caption{Dependence of the field of complete spin polarization on electron density at a temperature of 30~mK for two SiGe/Si/SiGe samples (dots and squares). The dashed black line is a linear fit to the high-density data which extrapolates to zero at $n_{\text s}\approx 0.14\times 10^{15}$~m$^{-2}$. The solid red line corresponds to the calculation \cite{fleury2010energy} for the clean limit. Top inset: the low density region of the main figure on an expanded scale. Also shown by the dotted red line is the calculation \cite{fleury2010energy} taking into account the electron scattering. Bottom inset: the parallel-field magnetoresistance at a temperature of 30~mK at different electron densities indicated in units of $10^{15}$~m$^{-2}$. The polarization field $B_{\text c}$ determined by the crossing of the tangents is marked by arrows. From Ref.~\cite{melnikov2017indication}.}
\label{fig2}
\end{figure}

\begin{figure}
\centering
\includegraphics[width=.9\linewidth]{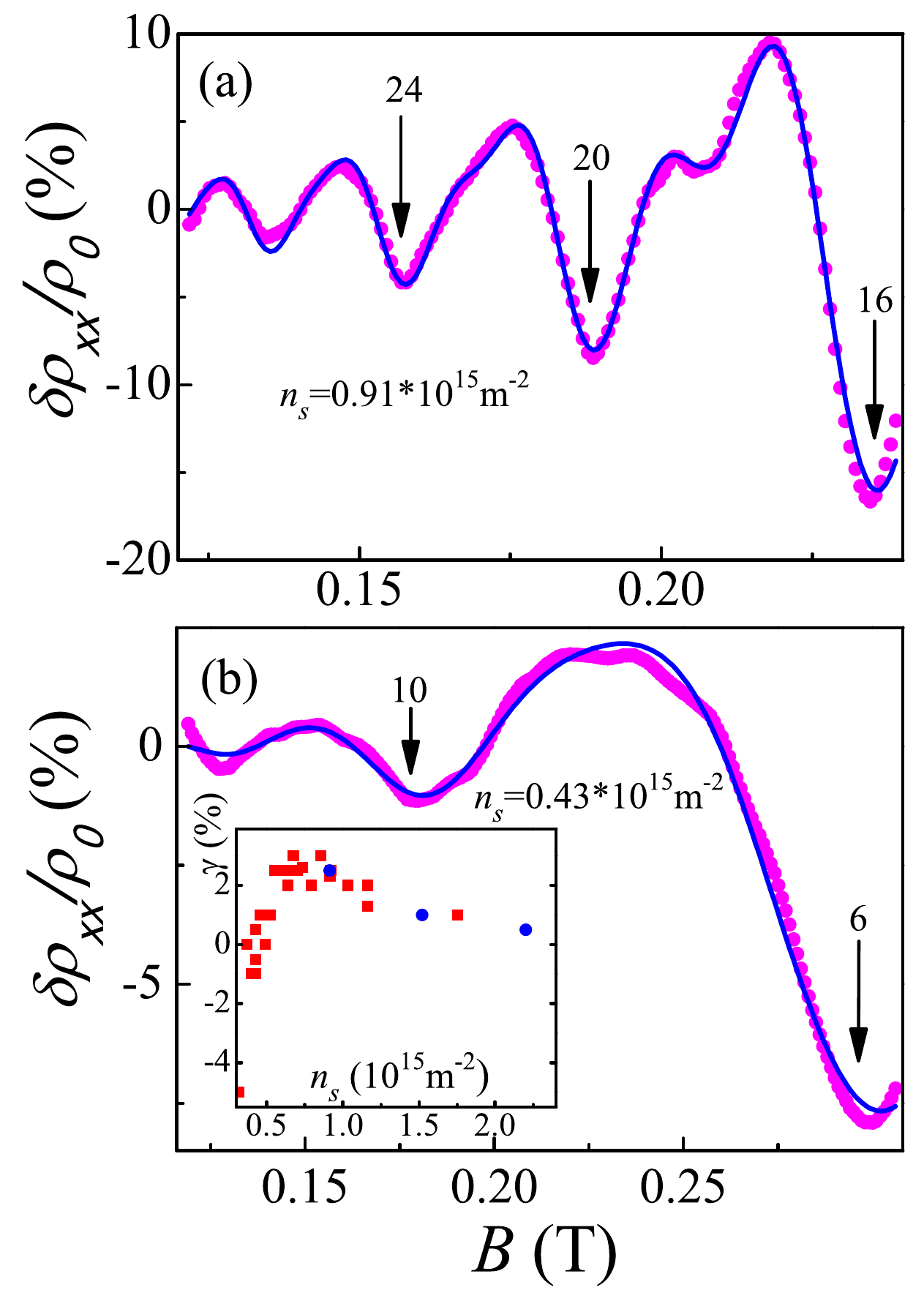}
\caption{Fits of the normalized magnetoresistance $\delta\rho_{\text{xx}}/\rho_0$ in a SiGe/Si/SiGe quantum well at a temperature of $\approx 30$~mK (dots) using Eq.~(\ref{S}) with (a) $g_{\text F}m_{\text F}/m_{\text e}=0.905$, $T_{\text D}=0.12$~K, $m_{\text F}=0.25\,m_{\text e}$, and $\gamma=2.5$\% and (b) $g_{\text F}m_{\text F}/m_{\text e}=1.11$, $T_{\text D}=0.15$~K, $m_{\text F}=0.33\,m_{\text e}$, and $\gamma=-0.5$\%. The filling factors $\nu=n_{\text s}hc/eB_\perp$ at minima are indicated. Inset: the asymmetry coefficient $\gamma$ \cite{pudalov2014probing} that describes the slightly different Dingle temperatures $T^{u,d}_D=T_D(1\pm\gamma)$ for two spin subbands versus electron density for two samples. The Dingle temperature for energetically favorable spin direction is smaller over the range of electron densities $0.6\times 10^{15}$~m$^{-2}<n_s<2\times 10^{15}$~m$^{-2}$, whereas at lower densities the quantity $\gamma$ changes sign. From Ref.~\cite{melnikov2017indication}.}
\label{fig3}
\end{figure}

To check whether or not the residual disorder affects the results for the magnetic field of complete spin polarization, we compare our data with those previously obtained on Si/SiGe samples with electron mobility an order of magnitude lower than that in our samples \cite{lu2008in}. At high electron densities, the dependence $B_{\text c}(n_{\text s})$ in Ref.~\cite{lu2008in} is also linear and extrapolates to zero at a finite density.  Furthermore, the slope of the dependence is equal to $6\times 10^{-15}$~T$\cdot$m$^2$ and is close to the slope $5.4\times 10^{-15}$~T$\cdot$m$^2$ observed in our experiment. However, the offset of approximately $0.3\times 10^{15}$~m$^{-2}$ in Ref.~\cite{lu2008in} is appreciably higher than that in our case. Therefore, the behavior of the polarization field $B_{\text c}$ is affected by the disorder potential in agreement with Refs.~\cite{fleury2010energy,renard2015valley}. A good agreement between our experimental data for $B_{\text c}$ and the calculations for the clean limit \cite{fleury2010energy} provides evidence that the electron properties of our samples are only weakly sensitive to the residual disorder, and the clean limit has been reached in our samples.

The product $g_{\text F}m$ that characterizes the whole 2D electron system can be determined in the clean limit from the equality of the Zeeman splitting and the Fermi energy of a completely spin-polarized electron system
\begin{equation}
g_{\text F}\mu_{\text B} B_{\text c}=\frac{2\pi\hbar^2n_{\text s}}{mg_{\text v}},\label{gm}
\end{equation}
where $g_{\text v}=2$ is the valley degeneracy and $\mu_{\text B}$ is the Bohr magneton.

On the other hand, the Land\'e $g$-factor $g_{\text F}$ and effective mass $m_{\text F}$ \textit{at the Fermi level} can be determined by the analysis of the Shubnikov-de Haas oscillations in relatively weak magnetic fields, as it was done in Ref.~\cite{melnikov2017indication}:
\begin{eqnarray}
A&=&\sum_iA^{LK}_i\cos\left[\pi i\left(\frac{\hbar c\pi n_{\text s}}{eB_\perp}-1\right)\right]Z^s_iZ^v_i\nonumber\\
A^{LK}_i&=&4\exp\left(-\frac{2\pi^2ik_{\text B}T_{\text D}}{\hbar\omega_{\text c}}\right)\frac{2\pi^2ik_{\text B}T/\hbar\omega_{\text c}}{\sinh\left(2\pi^2ik_{\text B}T/\hbar\omega_{\text c}\right)}\nonumber\\
Z^s_i&=&\cos\left(\pi i\frac{\Delta_{\text Z}}{\hbar\omega_{\text c}}\right)=\cos\left(\pi i\frac{g_{\text F}m_{\text F}}{2m_{\text e}}\right)\nonumber\\
Z^v_i&=&\cos\left(\pi i\frac{\Delta_{\text v}}{\hbar\omega_{\text c}}\right),\label{S}
\end{eqnarray}
where $T_{\text D}$ is the Dingle temperature, $T$ is the temperature, $m_{\text e}$ is the free electron mass, $\hbar\omega_{\text c}$ is the cyclotron splitting, $\Delta_{\text Z}$ is the Zeeman splitting, and $\Delta_{\text v}$ is the valley splitting. It is clear from Eq.~(\ref{S}) that as long as one sets $Z^v_i=1$ in the range of magnetic fields studied, the fitting parameters are $T_{\text D}m_{\text F}$, $m_{\text F}$, and $g_{\text F}m_{\text F}$ \cite{pudalov2014probing}. The values $T_{\text D}m_{\text F}$ and $m_{\text F}$ are obtained in the temperature range where the spin splitting is insignificant. Being weakly sensitive to these two fitting parameters, the shape of the fits at the lowest temperatures turns out to be very sensitive to the product $g_{\text F}m_{\text F}$. The quality of the fits is demonstrated in Fig.~\ref{fig3}. The magnetoresistance $\delta\rho_{\text xx}=\rho_{\text xx}-\rho_0$ normalized by $\rho_0$ (where $\rho_0$ is the monotonic change of the dissipative resistivity with magnetic field) is described well using Eq.~(\ref{S}).

The main result shown in Fig.~\ref{fig1} is that the products of the average $g_{\text F}m$ and $g_{\text F}m_{\text F}$ at the Fermi level behave similarly at high electron densities, where electron-electron interactions are relatively weak, but differ at low densities, where the interactions become especially strong \cite{dolgopolov2019two}. The product $g_{\text F}m_{\text F}$ monotonically increases as the electron density is decreased in the entire range of electron densities, while the product $g_{\text F}m$ saturates at low $n_{\text s}$. We emphasize that it is the qualitative difference in the behaviors of the two sets of data that matters, rather than a comparison of the absolute values. Taking into account the negligibility of the exchange effects in the 2D electron system in silicon \cite{kravchenko2004metal,shashkin2005metal}, this difference can only be attributed to the different behaviors of the two effective masses. Their qualitatively different behavior indicates the interaction-induced band flattening at the Fermi level in this electron system. To add confidence in our results and conclusions, we show in bottom inset in Fig.~\ref{fig1} the data for the effective mass $m_{\text F}$ determined by the analysis of the temperature dependence of the amplitude of Shubnikov-de~Haas oscillations, similar to Ref.~\cite{melnikov2014effective}. The similar behavior of $m_{\text F}$ and $g_{\text F}m_{\text F}$ with electron density allows one to exclude any possible influence of the $g$-factor on the behavior of the product of the effective mass and $g$-factor, which is consistent with the previously obtained results for the 2D electron system in silicon.

The experimental results are naturally interpreted within the concept of the fermion condensation
\cite{khodel1990superfluidity,nozieres1992properties,zverev2012microscopic} that occurs at the Fermi level in a range of momenta, unlike the condensation of bosons. With increasing strength of electron-electron interactions, the single-particle spectrum flattens in a region $\Delta p$ near the Fermi momentum $p_{\text F}$ (top inset to Fig.~\ref{fig1}). At relatively high electron densities $n_{\text s}>0.7\times 10^{15}$~m$^{-2}$, this effect is not important since the single-particle spectrum does not change noticeably in the interval $\Delta p$ and the behaviors of the energy-averaged effective mass and that at the Fermi level are practically the same. Decreasing the electron density in the range $n_{\text s}<0.7\times 10^{15}$~m$^{-2}$ gives rise to the flattening of the spectrum so that the effective mass at the Fermi level, $m_{\text F}=p_{\text F}/v_{\text F}$, continues to increase (here $v_{\text F}$ is the Fermi velocity). In contrast, the energy-averaged effective mass does not, being not particularly sensitive to this flattening. In the critical region, where the effective mass at the Fermi level tends to diverge, $m_{\text F}$ is expected to be temperature dependent. A weak decrease of the value $g_{\text F}m_{\text F}$ with temperature is indeed observed at the lowest-density point in Fig.~\ref{fig1}. In the critical region, the increase of $m_{\text F}$ is restricted by the limiting value determined by temperature: $m_{\text F} < p_{\text F}\Delta p/4k_{\text B}T$. In our experiments, the increase of $m_{\text F}$ reaches a factor of about two at $n_{\text s}=0.3\times 10^{15}$~m$^{-2}$ and $T\approx 30$~mK, which allows one to estimate the ratio $\Delta p/p_{\text F}\sim 0.06$. It is the smallness of the interval $\Delta p$ that provides good agreement between the calculation \cite{fleury2010energy} and our experiment.

It is worth noting that the effective mass at the Fermi level tends to diverge at a density $n_{\text m}$ higher than the critical electron density $n_{\text c}$ of the metal-insulator transition, revealing the qualitative difference between the ultralow-disorder SiGe/Si/SiGe quantum wells and the least-disordered Si MOSFETs where the opposite relation $n_{\text c}\ge n_{\text m}$ is found \cite{melnikov2019quantum}. This indicates that these two densities are not directly related, and the fermion condensation and metal-insulator transition are two different transitions.

\section{Merging of Landau levels in a strongly-interacting two-dimensional electron system}

Another example of a nontrivial manifestation of fermion interactions in strongly correlated Fermi liquids is the merging of quantum levels in a Fermi system with a discrete spectrum, in which case the fillings of the two quantum levels at the chemical potential are different \cite{shashkin2014merging}.

Application of the perpendicular magnetic field $B$ on a homogeneous 2D electron system creates two subsystems of Landau levels numbered $i$ and distinguished by $\pm$ projections of the electron spin on the field direction. The energy levels $\varepsilon_i^\pm$ in each set are spaced by the cyclotron splitting $\hbar\omega_{\text c}=\hbar eB/m^*c$, and the two sets of the Landau levels are shifted with respect to each other by the spin splitting $\Delta_{\text Z}=g\mu_{\text B}B$, where $m^*$ and $g$ are the values of mass and Land\'e $g$-factor renormalized by electron interactions (for simplicity, the valley degeneracy is so far neglected). The Landau levels with opposite spin directions should intersect with changing electron density, as caused by the strong dependence of the effective mass on $n_{\text s}$, provided the $g$-factor depends weakly on $n_{\text s}$. In particular, at high electron densities, the cyclotron splitting usually exceeds the spin splitting, whereas at low densities, the opposite case $\hbar\omega_{\text c}<\Delta_{\text Z}$ should occur due to the sharply increasing mass.

Both the thermodynamic and kinetic properties of the electron system are determined by the position of the chemical potential relative to the quantum levels, which is in turn determined by the magnetic field and electron density. The filling factor is equal to $\nu=n_{\text s}/n_0$, where $n_0=eB/hc$ is the level degeneracy. When $\nu$ is fractional, the chemical potential is pinned to the partially filled quantum level. The probability of finding an electron at the chemical potential is given by the fractional part of the filling factor and can be varied between zero and one. At the integer filling factor, there is a jump of the chemical potential. In an experiment, the jump manifests itself as a minimum in the longitudinal electrical resistance. The resistance minima in the $(B,n_{\text s})$ plane correspond to a Landau level fan chart.

\begin{figure}
\includegraphics[width=.9\linewidth]{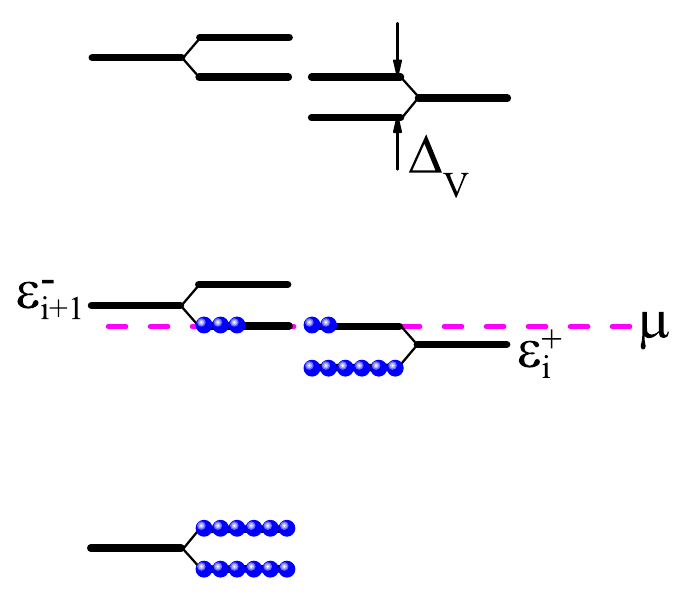}
\caption{Schematic diagram of merging of the spin- and valley-split Landau levels at the chemical potential. The occupied levels are indicated by dots. The fillings of the two quantum levels at the chemical potential vary with changing electron density. From Ref.~\cite{shashkin2014merging}.}
\label{fig4}
\end{figure}

\begin{figure}
\includegraphics[width=.9\linewidth]{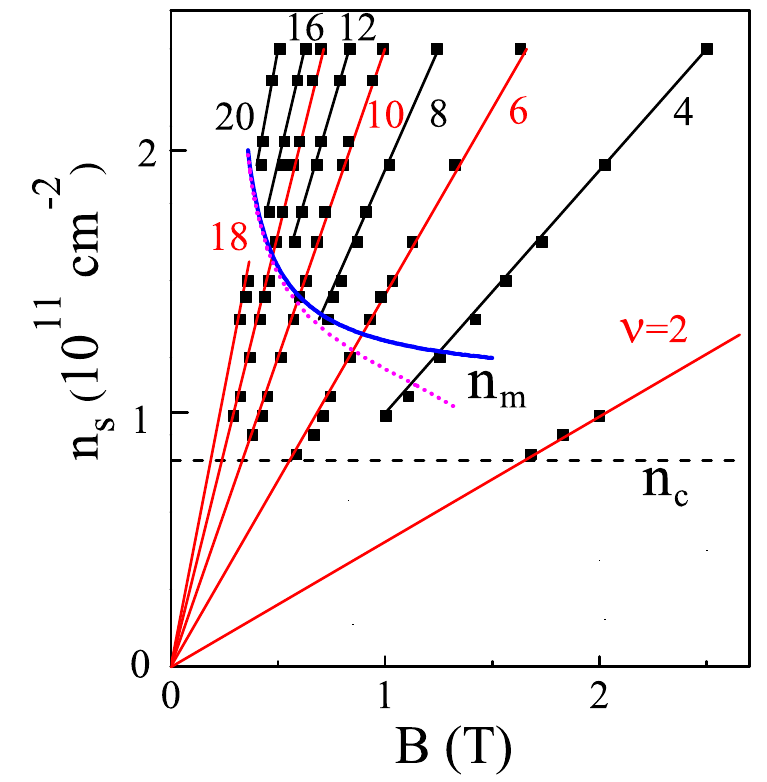}
\caption{Positions of the Shubnikov-de~Haas oscillation minima in a Si MOSFET sample in the ($B,n_{\text s}$) plane (squares) and the expected positions of the cyclotron and spin minima calculated according to the formula $n_{\text s}=\nu eB/hc$ (solid lines). The position of the metal-insulator transition in $B=0$ is indicated. The calculated merging boundary $n_{\text m}(B)$ is shown by the solid blue line if one uses Eq.~(\ref{upper}) and the dotted violet line if one takes into account nonlinear (cubic) corrections to the spectrum near the Fermi surface. From Ref.~\cite{shashkin2014merging}.}
\label{fig5}
\end{figure}

Provided that the external magnetic field is fixed and many quantum levels are occupied, the variation of the electron density in a quantum level is small compared to $n_{\text s}$. The variation of the energy $\varepsilon_\lambda$ is evaluated using the Landau relation
\begin{equation}
\delta\varepsilon_\lambda=\sum_\sigma\Gamma_{\lambda\sigma}\delta n_\sigma,\label{A}
\end{equation}
where $\Gamma_{\lambda\sigma}$ is the electron-electron interaction amplitude that is a phenomenological ingredient of the Fermi liquid theory \cite{landau1980statistical}. Selecting the magnetic field at which the difference between the neighboring Landau levels $\varepsilon_i^+$ and $\varepsilon_{i+1}^-$
\begin{equation}
D=\varepsilon_i^+-\varepsilon_{i+1}^-=\Delta_{\text Z}(n_{\text s},B)-\hbar\omega_{\text c}(n_{\text s},B)\label{dis}
\end{equation}
zeroes at the filling factor $\nu=n_{\text s}hc/eB=2i+2$, one starts from the higher density where both levels $(i+1)^-$ and $i^+$ are completely filled at $\nu=N=2i+3$, the difference $D(N)$ being negative. Removing the electrons from the level $(i+1)^-$ implies that the electron density decreases and $D$ increases. The level crossing occurs at $\nu=2i+2$, \textit{i.e.}, the level $i^+$ becomes empty and the level $(i+1)^-$ is completely filled, under the condition $\Gamma(i)=(\Gamma_{i+1,i+1}^{--}-\Gamma_{i,i+1}^{+-})-(\Gamma_{i+1,i}^{-+}-\Gamma_{i,i}^{++})\leq0$.

In the opposite case
\beq
\Gamma(i)>0,\label{ncm}
\eeq
the single-particle levels attract to each other and merge at the chemical potential $\mu$, as described by the merging equation $\varepsilon_{i+1}^-=\varepsilon_i^+=\mu$. Both levels exhibit partial occupation with fractions of empty states $0<f_i<1$ and $0<f_{i+1}<1$ that obey the normalization condition $f_i+f_{i+1}=f=N-\nu$. The merging starts when the empty states appear in the level $\varepsilon_i^+$ and ends when this level is completely emptied. This corresponds to the increase of the fraction of empty states $f$ in the range between $f=1$ (or $\nu=2i+2$) and $f=\min(1+\Gamma(i)/(\Gamma_{i+1,i+1}^{--}-\Gamma_{i,i+1}^{+-}),2)$. Outside the merging region, the conventional Landau level diagram is realized. Note that the gap between the neighboring Landau levels $\varepsilon_i^+$ and $\varepsilon_{i+1}^-$ proves to be invisible in transport and thermodynamic experiments. The upper boundary of the merging region $n_{\text m}(B)$ can be written
\beq
\hbar\omega_{\text c}-\Delta_{\text Z}=0.\label{upper1}
\eeq

The electron system in (100) Si MOSFETs is characterized by the presence of two valleys in the spectrum so that each energy level $\varepsilon_i^\pm$ is split into two levels, as shown schematically in Fig.~\ref{fig4}. One can easily see that the valley splitting $\Delta_{\text v}$ promotes the merging of quantum levels. The bigger the valley splitting, the higher the electron density at which the levels $(i+1)^-$ and $i^+$ with different valley indices should merge at the chemical potential at filling factor $\nu=4i+4$. The upper boundary of the merging region $n_{\text m}(B)$ is determined by the relation
\beq
\hbar\omega_{\text c}-\Delta_{\text Z}-\Delta_{\text v}=0\label{upper}
\eeq
that is different from Eq.~(\ref{upper1}) by the presence of the valley splitting. Since the electron density distributions corresponding to two valleys are spaced by distance $\alpha$ in the direction perpendicular to the Si-SiO$_2$ interface, the intervalley charge transfer creates an incremental electric field leading to $\Gamma(i)=4\pi e^2\alpha/\kappa$, where $\kappa$ is the dielectric constant. $\Gamma(i)$ determines the strength of the merging effect, and the lower boundary of the merging region of the neighboring Landau levels is given by the expression $\hbar\omega_{\text c}+n_0\Gamma(i)-\Delta_{\text Z}-\Delta_{\text v}=0$.

In the high-density limit, where the effects of electron-electron interactions are negligible, the effective mass and $g$ factor are equal to $m_0=0.19\, m_{\text e}$ and $g_0=2$ so that the cyclotron splitting significantly exceeds the spin splitting. At low electron densities, where the interaction effects are strong, the effective mass $m^*(n_{\text s})$ is found to diverge as $m_0/m^*\simeq(n_{\text s}-n_{\text c})/n_{\text s}$ at the quantum critical point close to the metal-insulator transition which occurs at $n_{\text c}\simeq 8\times 10^{10}$~cm$^{-2}$, while the $g$-factor stays close to $g_0$, being equal to $g\simeq 1.4\, g_0$ \cite{shashkin2002sharp,kravchenko2004metal,shashkin2005metal,mokashi2012critical}. The Landau level fan diagram for this electron system in perpendicular magnetic fields is represented in Fig.~\ref{fig5}. The quantum oscillation minima at filling factor $\nu=4i+4$ disappear below some electron density $n^*$ depending on $\nu$, while the minima at $\nu=4i+2$ persist down to appreciably lower densities. Although this behavior is consistent with the sharp increase of the effective mass with decreasing $n_{\text s}$, the dependence of the density $n^*$ on the filling factor (or $B$) turns out to be anomalously strong and lacks explanation. Particularly, this cannot be accounted for by the impurity broadening of quantum levels in terms of $\omega_{\text c}\tau\sim1$ (where $\tau$ is the elastic scattering time) in which case the drop of mobility $e\tau/m^*$ at low electron densities is controlled by the increasing mass \cite{shashkin2002sharp}.

The expected upper boundary of the merging region $n_{\text m}(B)$, shown by the solid blue line in Fig.~\ref{fig5}, has been determined in Ref.~\cite{shashkin2014merging}. The calculated boundary is in agreement with the experimental density $n^*(B)$ at which the oscillation minima at $\nu=4i+4$ vanish. This fact gives evidence for the level merging in a 2D electron system in silicon.

The description of the high-field data $n^*(B)$ improves within the merging picture if one takes into account nonlinear (cubic) corrections to the spectrum at the Fermi surface near the quantum critical point that naturally lead to a decrease of the effective mass with the magnetic field. The corrected dependence $n_{\text m}(B)$ is shown by the dotted violet line in Fig.~\ref{fig5}; for more on this, see Ref.~\cite{shashkin2014merging}.

\section{Interaction-induced merging of Landau levels in an electron system of double quantum wells}

The spectrum of a two-dimensional electron system subjected to a perpendicular magnetic field consists of two equidistant ladders of quantum levels for the spin up and down directions, as considered in the preceding section. If the magnetic field is tilted by an angle $\beta$, the spacing between the quantum levels in each of the spin ladders is equal to $\hbar\omega_{\text c}=\hbar eB\cos(\beta)/m^*c$, and the shift between the ladders equals $g\mu_{\text B}B$. Increasing the tilt angle leads to the crossing of the quantum levels of the two ladders. The crossing happens for the first time at an angle $\beta_1$ that satisfies the condition $\cos(\beta_1)=gm^*/2m_e$. At $\beta=\beta_1$, the chemical potential jumps at even filling factors and the corresponding fan chart lines should
disappear. If one takes into account the interaction between the electrons of neighboring quantum levels and increases the tilt angle in the vicinity of $\beta_1$, then, tentatively, the quantum level filled before crossing should have got emptied with increasing $\beta$. However, suppose the single-particle energy of electrons on the emptying level decreases due to the electron interaction. In that case, both levels remain pinned to the chemical potential over a wide range of angles $\Delta\beta_1$ that is determined by the interaction strength. The probability of finding an electron at the chemical potential is different for opposite spin orientations, depending on the external parameter, the tilt angle. Such a behavior corresponds to the merging of quantum levels.

\begin{figure}
\includegraphics[width=.9\linewidth]{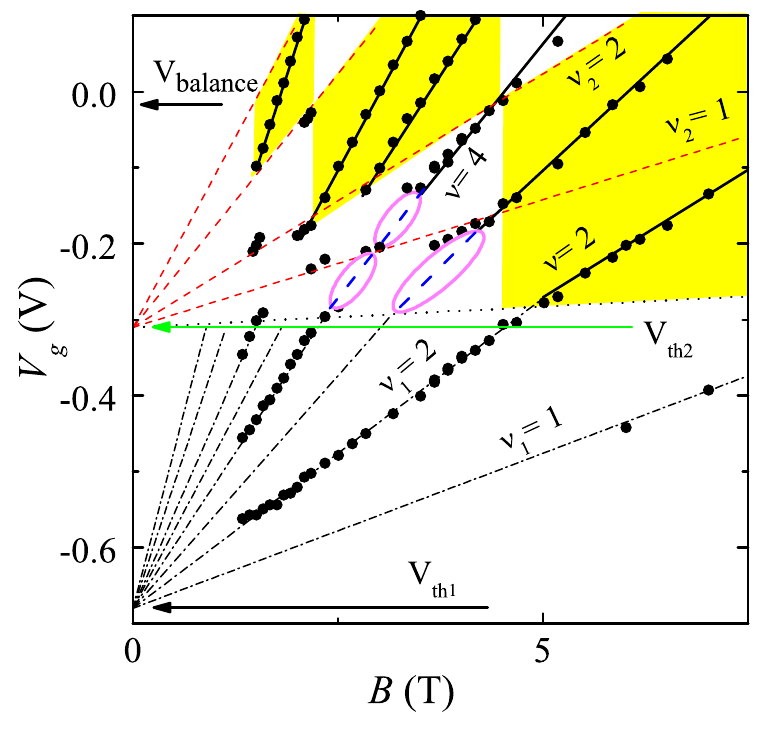}
\caption{Landau level fan chart for the AlGaAs double quantum well studied in Ref.~\cite{shashkin2015interaction}. Positions of the longitudinal resistance minima in the $(B,V_{\text g})$ plane are marked by the dots. The filling factor $\nu$ for the double layer electron system as well as the filling factor $\nu_1$ ($\nu_2$) for the back (front) layer are indicated. Over the shaded areas, the merging of quantum levels in perpendicular magnetic fields is impossible. In the regions marked by the ovals, no resistance minima are observed in a perpendicular magnetic field, whereas these appear in a tilted magnetic field. From Ref.~\cite{shashkin2015interaction}.}
\label{fig6}
\end{figure}

In the above hypothetical consideration, the crossing or merging of quantum levels is controlled by the tilt angle of the magnetic field. In the experiments on a strongly-interacting 2D electron system in (100) Si MOSFETs, the disappearance of the longitudinal resistance minima is analyzed when changing both the perpendicular magnetic field and electron density at fixed filling factor $\nu=4(i+1)$, where $i$ is an integer. In this case, the level merging occurs near the quantum critical point, as controlled by the effective mass depending on electron density \cite{shashkin2014merging}. One might think that the level merging is a precursor of the Fermi surface swelling. In fact, the two effects are not necessarily related to each other. Below it is demonstrated that the effect of level merging occurs in a bilayer 2D electron system with a tunnel barrier between the electron layers \cite{shashkin2015interaction}. Note that the effective mass enhancement is insignificant in this case.

The sample used in this section is a parabolic quantum well with a narrow tunnel barrier grown on a GaAs substrate (for a detailed description, see Ref.~\cite{shashkin2015interaction}). Applying a voltage $V_{\text g}$ between the gate and the contact to the quantum well makes it possible to tune the electron density. The electrons appear in the back part of the quantum well when the gate voltage is above $V_{\text{th1}}\approx -0.7$~V and occupy one subband up to $V_{\text g}=V_{\text{th2}}\approx -0.3$~V (Fig.~\ref{fig6}). At $V_{\text g}> V_{\text{th2}}$, the electrons appear in the front part of the well and fill the second subband up to the balance point $V_{\text g}=V_{\text{balance}}\approx 0$.

The authors focus on the range of gate voltages between $V_{\text{th2}}$ and $V_{\text{balance}}$ where the electrons in perpendicular magnetic fields occupy the two quantum ladders. Positions of the quantum levels are determined by the magnetic field and gate voltage. Over the shaded areas in Fig.~\ref{fig6}, the gaps in the single-particle spectrum and the chemical potential jumps are protected by quantum effects \cite{dolgopolov1999electron}. In the remaining areas, the merging of quantum levels is, in principle, possible.

\begin{figure}
\includegraphics[width=.9\linewidth]{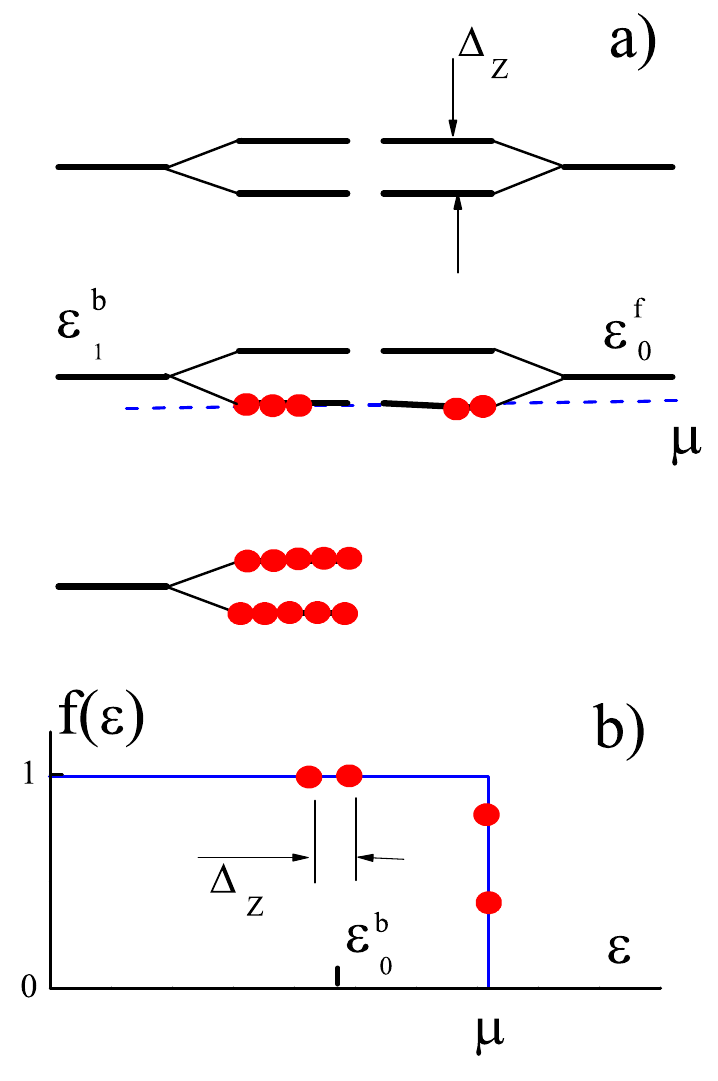}
\caption{(a) Layout and filling of the quantum levels in the bilayer electron system in the merging regime at filling factor $\nu=3$. (b)~The distribution function of the electrons in the merging regime at $\nu=3$. From Ref.~\cite{shashkin2015interaction}.}
\label{fig7}
\end{figure}

Below, the filling factor $\nu=3$ is considered. At $V_{\text g}=V_{\text{th2}}$, the magnetic field is equal to $B_\nu=n_1(V_{\text{th2}})hc/3e$, where $n_1(V_{\text{th2}})$ is the electron density in the back layer. The energy $\varepsilon_0-g\mu_{\text B}B_\nu/2$ of the spin up level of the front layer ladder is the same as the energy $\varepsilon_1-g\mu_{\text B}B_\nu/2$ of the spin up level of the back layer ladder (Fig.~\ref{fig7}(a)). Since far from the balance point, the electron density in the back layer remains practically unchanged with increasing $V_{\text g}$ above $V_{\text{th2}}$ (see, \textit{e.g.}, Fig.~6(b) of Ref.~\cite{davies1996hybridization}), the electron density in the front layer in a magnetic field $B=B_\nu+\Delta B$ along the dashed line at $\nu=3$ restricted by the oval in Fig.~\ref{fig6} is equal to $n_2\simeq\Delta n=3e\Delta B/hc$. To balance the change in the cyclotron energy $\hbar e\Delta B/m^*c$ and have both levels pinned to the chemical potential $\mu$, it is necessary to transfer a small amount of electrons between the levels
\begin{equation}
\delta n_2=-\delta n_1>0,\label{eq1}
\end{equation}
which gives rise to a shift of the single-particle levels
\begin{eqnarray}
\delta\varepsilon_1^b&=&\Gamma_1^{bb}\delta n_1+\Gamma_1^{bf}\delta n_2 \nonumber\\
\delta\varepsilon_0^f&=&\Gamma_0^{fb}\delta n_1+\Gamma_0^{ff}\delta n_2,\label{eq2}
\end{eqnarray}
where $\Gamma_j^{\lambda\sigma}$ is the electron interaction amplitude, the index $b$ ($f$) refers to the back (front) layer, and $j=0,1$ is the Landau level number. Both levels are pinned to the chemical potential under the condition $\delta\varepsilon_0-\delta\varepsilon_1=\hbar e\Delta B/m^*c$, which yields
\begin{equation}
\delta n_2=\frac{\hbar e\Delta B}{m^*c\,\Gamma},\label{eq3}
\end{equation}
where $\Gamma=\Gamma_1^{bb}+\Gamma_0^{ff}-\Gamma_1^{bf}-\Gamma_0^{fb}$. In the parallel-plate-capacitor approximation, one gets
\begin{equation}
\Gamma\simeq\frac{4\pi e^2a}{\kappa},\label{eq4}
\end{equation}
where $a$ is the distance between the weight centers of the electron density distributions in the $z$ direction in both subbands. The level merging holds for the filling factor
\begin{equation}
\nu_2=\frac{(n_2+\delta n_2)hc}{eB}<1.\label{eq5}
\end{equation}

The authors stress that in the wide range of magnetic fields at fixed filling factor $\nu$, the probability of finding an electron with energy equal to the chemical potential is different for the two merged levels, as shown in Fig.~\ref{fig7}(b).

Although the case of filling factor $\nu=3$ has been considered above for simplicity, the same arguments are also valid for higher filling factors.

The occurrence of the merging of quantum levels in the experiment is confirmed by using tilted magnetic fields. With tilting magnetic field, the magnetoresistance minima and chemical potential jumps arise \cite{deviatov2000opening} particularly along the dashed lines at $\nu=3$ and $\nu=4$ indicated by the ovals in Fig.~\ref{fig6}. The appearance of the chemical potential jumps in the double layer electron system in tilted magnetic fields signals that the quantum levels are narrow enough.

As has been mentioned above, the chemical potential jumps can be protected by quantum effects. In general, a transfer of electrons between the quantum levels of different subbands leads to mixing the wave functions of the subbands and opening an energy gap if the non-diagonal matrix elements are not equal to zero \cite{dolgopolov1999electron}. This is realized over the shaded areas in Fig.~\ref{fig6}. In contrast, in the merging regions at $\nu=3$ and $\nu=4$ indicated by the ovals in Fig.~\ref{fig6}, the non-diagonal matrix elements in perpendicular magnetic fields are equal to zero because of the orthogonality of the in-plane
part of the wave functions in the bilayer electron system. Tilting the magnetic field breaks the orthogonality of the wave functions of the neighboring quantum levels, and the energy gap emerges \cite{deviatov2000opening,duarte2007landau}.

\section{Merging of the quantum levels of composite fermions}

Finally, we consider an indication of merging of the quantum levels of composite fermions with different valley indices \cite{dolgopolov2021valley}. The concept of composite fermions \cite{jain1989composite,halperin1993theory,jain1994composite,chakraborty2000electron,jain2007composite} can successfully describe the fractional quantum Hall effect with odd denominators by reducing it to the ordinary integer quantum Hall effect for composite particles. In the simplest case, the composite fermion consists of an electron and two magnetic flux quanta and moves in an effective magnetic field $B^*$ given by the difference between the external magnetic field $B$ and the field corresponding to the filling factor for electrons, equal to $\nu=1/2$. The filling factor for composite fermions, $p$, is connected to $\nu$ according to the expression $\nu=p/(2p\pm1)$. The fractional energy gap, which is predicted to be determined by the Coulomb interaction in the form $e^2/\kappa l_{\text B}$, corresponds to the cyclotron energy of composite fermions $\hbar\omega_{\text c}^*=\hbar eB^*/m_{\text{CF}}c$, where $l_{\text B}=(\hbar c/eB)^{1/2}$ is the magnetic length and $m_{\text{CF}}$ is the effective composite fermion mass. The electron-electron interactions enter the theory \cite{jain1989composite,halperin1993theory,jain1994composite,chakraborty2000electron,jain2007composite} implicitly because a mean-field approximation is employed, assuming that the electron density fluctuations are small. The theory is confirmed by the experimental observation of a scale corresponding to the Fermi momentum of composite fermions in zero effective magnetic field at $\nu=1/2$.

\begin{figure}
\scalebox{0.42}{\includegraphics[angle=0]{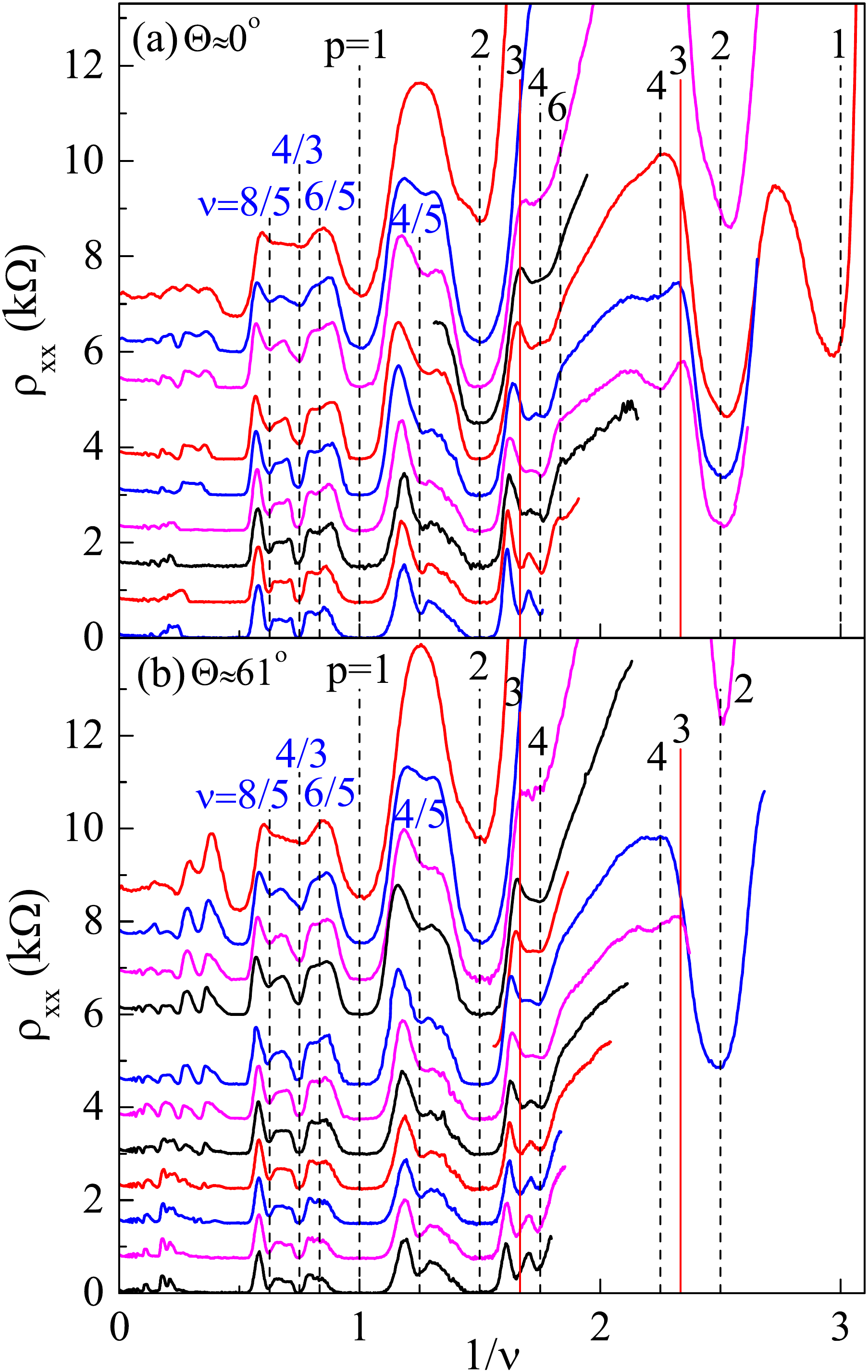}}
\caption{Magnetoresistance in a SiGe/Si/SiGe quantum well at $T\approx 0.03$~K (a)~in perpendicular magnetic fields at electron densities (from top to bottom) 2.14, 2.81, 3.48, 3.81, 4.15, 4.82, 5.49, 6.15, 6.82, and $7.49\times 10^{10}$~cm$^{-2}$, and (b)~in tilted magnetic fields at electron densities (from top to bottom) 2.14, 2.81, 3.48, 4.15, 4.48, 4.82, 5.49, 6.15, 6.82, 7.49, 8.16, and $8.83\times 10^{10}$~cm$^{-2}$. Curves are vertically shifted by 750~$\Omega$ for clarity. Dashed vertical lines mark the expected positions of the observed minima of the resistance, and solid vertical lines correspond to the minima that are expected, but not observed at low densities. From Ref.~\cite{dolgopolov2021valley}.}
\label{fig8}
\end{figure}

Samples studied in this section are ultraclean bivalley (001) SiGe/Si/SiGe quantum wells similar to those described in Refs.~\cite{melnikov2015ultra,melnikov2017unusual}. The longitudinal resistivity $\rho_{\text {xx}}$ as a function of the inverse filling factor is shown for different electron densities in Fig.~\ref{fig8}(a). The resistance minima are seen at composite fermion quantum numbers $p=1$, 2, 3, 4, and 6 near $\nu=1/2$ in positive and negative effective fields $B^*$, revealing the high quality of the sample. The high quality of the quantum well is also confirmed by the presence of the $\nu=4/5$ and $\nu=4/11$ fractions \cite{dolgopolov2018fractional}, corresponding to $p=4/3$, which can be described in terms of the second generation of composite fermions. The minima at $p=3$ disappear below a certain electron density, although the surrounding minima at $p=2$ and $p=4$ persist to significantly lower densities. Clearly, the prominence of the minima at $p=3$ at low electron densities cannot be explained by level broadening. On the other hand, this finding is strikingly similar to the effect of the disappearance of the cyclotron minima in the magnetoresistance at low electron densities in Si MOSFETs while the spin minima survive down to appreciably lower densities \cite{kravchenko2000shubnikov}, which signifies that the cyclotron splitting becomes equal to the sum of the spin and valley splittings, and the corresponding valley sublevels merge \cite{shashkin2014merging}.

\begin{figure}
\scalebox{0.4}{\includegraphics[angle=0]{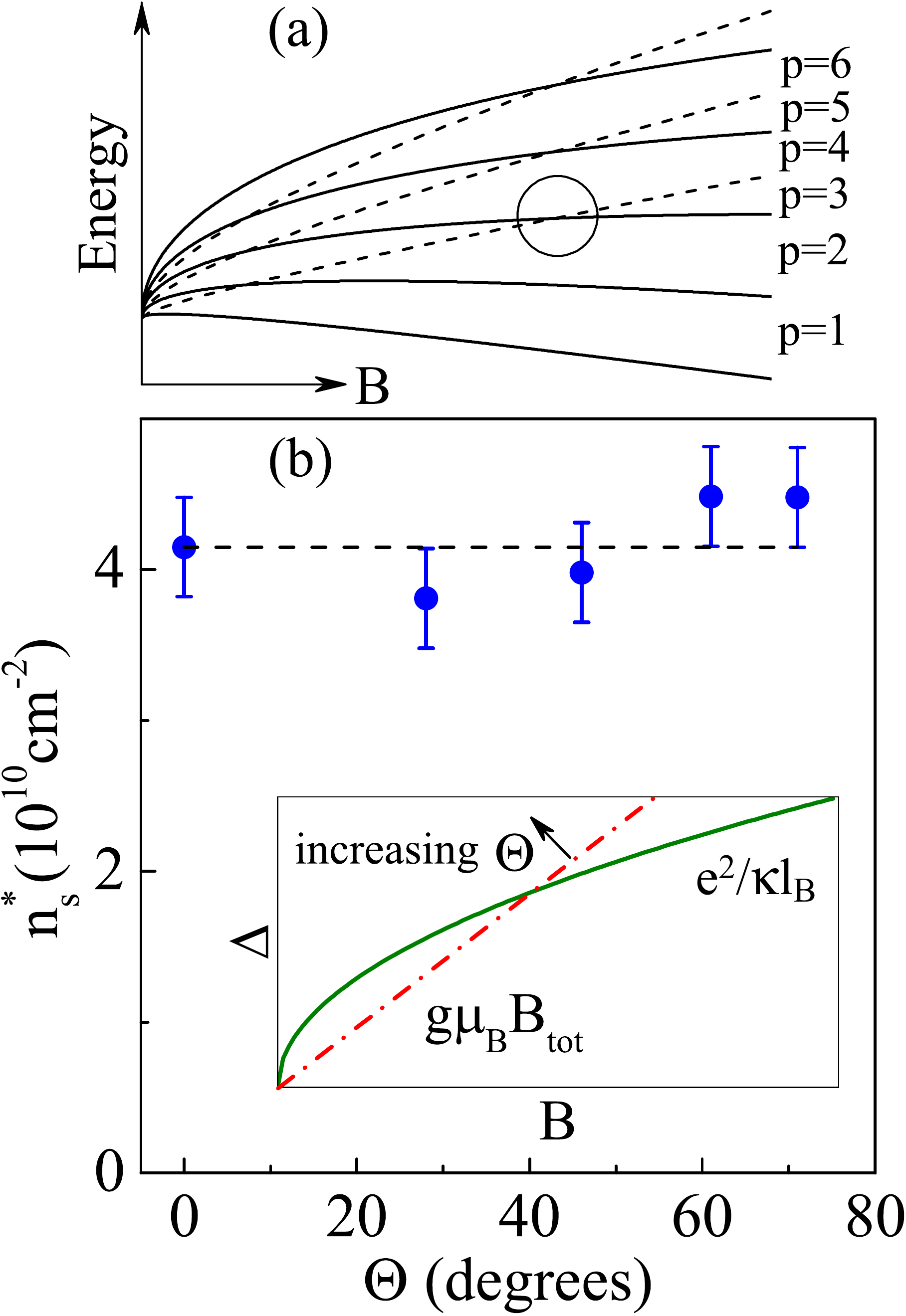}}
\caption{(a) The schematic behavior of the composite fermion levels taking account of the splitting between upper (dashed lines) and lower (solid lines) valleys with changing magnetic field $B$ at fixed $p$. In the region of interest at $p=3$, indicated by the circle, there occurs either a simple crossing of the levels or merging/locking of the levels accompanied with a gradual change in the fillings of both levels. (b) The onset density $n_{\text s}^*$ of the resistance minimum at $\nu=3/5$ in a SiGe/Si/SiGe quantum well as a function of the tilt angle. The dashed horizontal line is a fit to the data. The inset schematically (up to a numerical factor) shows the cyclotron energy of composite fermions (the solid line) and the Zeeman energy (the dotted-dashed line) as a function of the
magnetic field $B$ at a fixed tilt angle. The slope of the straight line increases with increasing $\Theta$. From Ref.~\cite{dolgopolov2021valley}.}
\label{fig9}
\end{figure}

Measurements in tilted magnetic fields allow one to distinguish between the spin and valley origin of the effect. The magnetoresistance as a function of the inverse filling factor is shown for the tilt angle $\Theta\approx 61^\circ$ at different electron densities in Fig.~\ref{fig8}(b). Here, the authors focus on the resistance minimum at $\nu=3/5$. The behavior observed for the $\nu=3/5$ minimum is very similar to that in perpendicular magnetic fields, which holds for all samples and tilt angles. One determines the onset $n_{\text s}^*$ for the $\nu=3/5$ minimum and plots it versus the tilt angle, as shown in Fig.~\ref{fig9}(b).  The value $n_{\text s}^*$ turns out to be independent, within the experimental uncertainty, of the tilt angle of the magnetic field. Since the spin splitting is determined by total magnetic field, $\Delta_{\text s}=g\mu_{\text B}B_{\text{tot}}$, one expects that the onset $n_{\text s}^*$ should decrease with the tilt angle [the inset to Fig.~\ref{fig9}(b)], which is in contradiction with the experiment. The authors conclude that the spin origin of the effect can be excluded, revealing its valley origin. The valley splitting $\Delta_{\text v}$ is expected to be insensitive to the parallel component of the magnetic field \cite{khrapai2003direct} so that the value $n_{\text s}^*$ should be independent of the tilt angle, which is consistent with the experiment. Thus, these results indicate the intersection or merging of the quantum levels of composite fermions with different valley indices, which reveals the valley effect on the fractions.

It is clear that for the occurrence of the crossing or merging of the levels of composite fermions with different valley indices, the functional dependences of both splittings on magnetic field (or electron density) at fixed $p$ should be different. Indeed, the cyclotron energy of composite fermions $\hbar\omega_{\text c}^*$ is determined by the Coulomb interaction energy $e^2/\kappa l_{\text B}$, and the valley splitting $\Delta_{\text v}$ in a 2D electron system in Si changes linearly with changing magnetic field (or electron density) \cite{khrapai2003strong}. In high magnetic fields, the valley splitting strongly exceeds the cyclotron energy of composite fermions so that for the case of $p=3$, all three filled levels of composite fermions belong to the same valley [Fig.~\ref{fig9}(a)]. As the magnetic field is decreased at fixed $p$, the lowest level with the opposite valley index should become coincident with the top filled level, leading to the vanishing of the energy gap and the disappearance of the resistance minimum at $p=3$. With a further decrease of the magnetic field there should occur either a simple crossing of the levels and reappearance of the gap or merging/locking of the levels accompanied by a gradual change in the fillings of both levels \cite{shashkin2014merging}. In analogy with Si MOSFETs, it is very likely that the merging of the composite fermion levels with different valley indices occurs in ultra-low-disorder SiGe/Si/SiGe quantum wells.

\section{Conclusions}

In conclusion, we have reviewed recent experimental results pointing to the band flattening and Landau level merging at the chemical potential in strongly correlated 2D electron systems. It is shown that the occupation numbers of quantum states at the chemical potential can be different within the range between zero and one, which reveals the non-Fermi-liquid form of the distribution function.

\section{Acknowledgments}

The ISSP group was supported by a Russian Government contract. S.V.K. was supported by NSF Grant No.\ 1904024.

%\bibliography{references}

%\end{document}

\end{document}